\documentclass{aastex}          
\usepackage{spr-astr-addons}    

\begin{document}
%
\title{GeV-TeV $\gamma$-ray energy spectral break of BL Lac objects}

\shorttitle{Short article title}
\shortauthors{<Zhong et al.>}


\author{W. Zhong.\altaffilmark{1}} \and 
\author{W.G. Liu.\altaffilmark{1}}
\and
\author{Y.G. zheng.\altaffilmark{1,2,3}}
\affil{ynzyg@ynu.edu.cn}
\altaffiltext{1}{Department of Physics, Yunnan Normal University, Kunming, 650092, China.}
\altaffiltext{2}{Yunnan Observatories, Chinese Academy of Sciences, Kunming 650011, China.}
\altaffiltext{3}{Key Laboratory for the Structure and Evolution of Celestial Objects, Chinese Academy of Sciences, Kunming 650011, China.}

\begin{abstract}
  In this paper, we compile the very-high-energy and high-energy spectral indices of 43 BL Lac objects from the literature. Based on a simple math model, $\Delta\Gamma_{obs}=\alpha {\rm{z}}+\beta $, we present evidence for the origin of an observed spectral break that is denoted by the difference between the observed very-high-energy and high-energy spectral indices, $\Delta\Gamma_{obs}$. We find by linear regression analysis that $\alpha\ne 0$ and $\beta\ne 0$. These results suggest that the extragalactic background light attenuation and the intrinsic curvature dominate on the GeV-TeV $\gamma$-ray energy spectral break of BL Lac objects. We argue that the extragalactic background light attenuation is an exclusive explanation for the redshift evolution of the observed spectral break.
\end{abstract}

\keywords{BL Lacertae objects: general; radiation mechanisms: non-thermal; methods: statistical.}

\section{Introduction}
BL Lac objects show that continuum emission, arising from the jet emission taking place in an AGN whose jet axis is closely aligned with the observer's line of the sight, is dominated by non-thermal emission, as well as rapid and large amplitude variability. Their spectral energy distributions (SEDs) are characterized by two distinct broad bumps, implying two emission components (e.g., \citealt{phdthesis2009}). It is widely believed that the first bump is produced by electron synchrotron radiation. The second bump is probably produced by inverse Compton scattering off the relativistic electrons, either on the synchrotron photons \citep{Maraschi1992} or on some other photon populations(\citealt{Dermer1993Model}; \citealt{Sikora1994Comptonization}). These processes should result in a smooth spectrum in the $\gamma$-ray band. However, the SED shows a tendency with steepening toward higher energies (\citealt{costamante2013gamma-rays}; \citealt{dwek2013the}). This tendency indicates a spectral break in different bands.

Extragalactic background light (EBL) is the diffuse and isotropic radiation fields from ultraviolet to far-infrared wavelengths \citep{Hauser2001THE}. It plays an important role in the formation and evolution of stellar objects and galaxies. Since the very-high-energy (VHE, 100 GeV$\le$ E $<$ 10TeV) $\gamma$-ray photons propagate through cosmic space, they can interact with EBL photons producing the electron-positron pairs (i.e., $\gamma_{VHE} + \gamma_{EBL}\to e^{+} + e^{-}$, e.g., \citealt{Stecker1992TeV}; \citealt{Stecker1998Absorption}; \citealt{Stecker2006Intergalactic}; \citealt{A2008Extragalactic}), which will steepen the observed spectrum (\citealt{Ackermann2012the}; \\\citealt{Sanchez2013Evidence}). If the $\gamma$-ray radiation from BL Lacs cannot be attenuated by either the secondary cascade (\citealt{Essey2012On}; \citealt{Zheng2013Evidence}; \citealt{Zheng2016Bethe}) or the axion-like particle conversion(\citealt{Horns2012Indications}), we should expect a distinctive EBL attenuation characteristic (e.g., \citealt{Kneiske2004Implications}; \citealt{Stecker2006Intergalactic}; \\ \citealt{Imran2008Detecting}; \citealt{A2008Extragalactic}; \citealt{Tavecchio2009Intrinsic}) at high energies in gamma-ray sources. Taking the EBL attenuation into account during the $\gamma$-ray propagation, we expect a difference in the observed high energy (HE, 100 MeV $\le$ E $<$ 300GeV) and VHE spectrum of BL Lacs.

In order to determine the spectral break, \\ \cite{Dom2015Spectral} selected 128 extragalactic sources from the second catalog of Fermi-LAT sources (2FHL), which is observed in HE bands extending in energy from 50 GeV to 2 TeV. Nevertheless, the study of the spectral break mechanism with the 2FHL catalog cannot explain the spectrum detected with energies higher than 2 TeV. Owing to the MAGIC, HESS, and VERITAS broadening of the 2FHL spectra (\citealt{Funk2012The}), we can study the relativistic particle acceleration (\citealt{Holder2012TeV}) and the spectral energy break. Additionally, the sources of the 2FHL catalog used by \cite{Dom2015Spectral} included the flat spectrum radio quasars (FSRQs), whose GeV-TeV spectrum would be contributed by the photon populations of the broad line region (BLR) or of the accretion disk (\citealt{Poutanen2010GeV}; \citealt{Ackermann2012the}). This results from without a significant EBL attenuation in the SED of FSRQs. Therefore, we focus on the spectra of BL Lacs.

In order to clarify the spectral break mechanism, in this paper we focus on the analysis of the GeV-TeV energy spectral index of the BL Lac objects. Our goal is to determine whether the EBL attenuation is the exclusive origin of the energy spectral break of the BL Lac objects. Since the observed spectrum is attenuated by EBL, we should obtain the absorption-corrected spectrum. In Section 2, we describe the sample; in Section 3, we compare the observed spectrum and intrinsic spectrum; in Section 4, we analyze the origin of spectral break; and in Section 5, we provide our conclusions and a discussion.

\section{Sample Description}
We compile the sources from the TeV catalog \footnote{http://tevcat.uchicago.edu} and Fermi LAT third source Catalog (3FGL)\footnote{http://www.asdc.asi.it/fermi3fgl/} to determine the spectral break. The TeV catalog provides the observed VHE spectral indices (\citealt{Holder2012TeV}), and the 3FGL Catalog provides the observed HE spectral indices. Since the intrinsic spectral indices (unattenuated by the EBL) cannot be obtained directly, we should correct to the observed VHE spectrum.

Generally, the observed VHE spectrum could be described by $dN/dE\propto E^{- \Gamma_{HE, obs}}$ and the observed VHE spectrum could be described by $dN/dE\propto E^{- \Gamma_{VHE, obs}}$, where $\Gamma_{HE, obs}$, $\Gamma_{VHE, obs}$ is the observed spectral index in the HE and VHE regimes, respectively. In most cases, the distinct attenuation signature on the spectrum appears above 100 GeV (\citealt{Zheng2013Evidence}; \citealt{Zheng2016Bethe}), so it indicates that the observed HE spectral index is equivalent to the intrinsic HE spectral index (i.e.,$\Gamma_{HE, obs}=\Gamma_{HE, int}$). Compared to the observed VHE spectrum, the intrinsic VHE spectrum is an EBL-corrected spectrum, which could be obtained by the following equation (\citealt{AlbertJ2007Observation}, \citealt{Raue2008OPTICAL}):
\begin{eqnarray}
\frac{{dN}}{{d{E_{{\mathop{\rm int}} }}}} = \frac{{dN}}{{d{E_{obs}}}} \ {e^{{\tau _{\gamma \gamma }}(E,z)}},
\label{Eq:1}
\end{eqnarray}

where ${\tau _{\gamma \gamma }}(E,z)$ is the EBL absorption depth for a photon with energy  $E$ traveling from a source at redshift $z$. We use the average EBL model of \cite{Dwek2005Simultaneous} to calculate ${\tau _{\gamma \gamma }}(E,z)$ and then obtain the intrinsic VHE spectral index.

Table 1 lists 43 observed and intrinsic VHE spectral indices (i.e., $\Gamma_{VHE, obs}$ and $\Gamma_{VHE, int}$) and 43 HE spectral indices (i.e., ${\Gamma _{HE}}$) that are associated with the VHE spectrum. According to the synchronous peak frequency, one of all sources is low-frequency-peaked BL Lacs (LBL, $\nu _{{\rm{peak}}}^S \le {10^{14}}Hz$), five sources are intermediate-frequency-peaked BL Lacs (IBL, ${10^{14}}Hz \le \nu _{{\rm{peak}}}^S \le {10^{15}}Hz$), and the other sources are high-frequency-peaked BL Lacs (HBL, $\nu _{{\rm{peak}}}^S \ge {10^{15}}Hz$).

\section{Comparing Observed Spectrum and Intrinsic Spectrum}

In Figure 1 we show the distribution of the intrinsic spectral indices and observed spectral indices. It can be seen that the distributions of intrinsic spectral indices are concentrated in the region of a smaller index. We use the likelihood methodology described by \cite{Venters2007The} to estimate the mean and spread of distributions. We find that the mean observed and intrinsic indices are $\Gamma _{0, obs}= 3.10$ and  $\Gamma _{0, int}= 2.05 $, respectively. These signatures are the certain results of EBL corrections. Since the value of the mean intrinsic index is less than the mean observed index, we thus believe that the intrinsic spectrum is a hard spectrum.

\begin{figure}[tb]
  \centering
  \includegraphics[width=8cm]{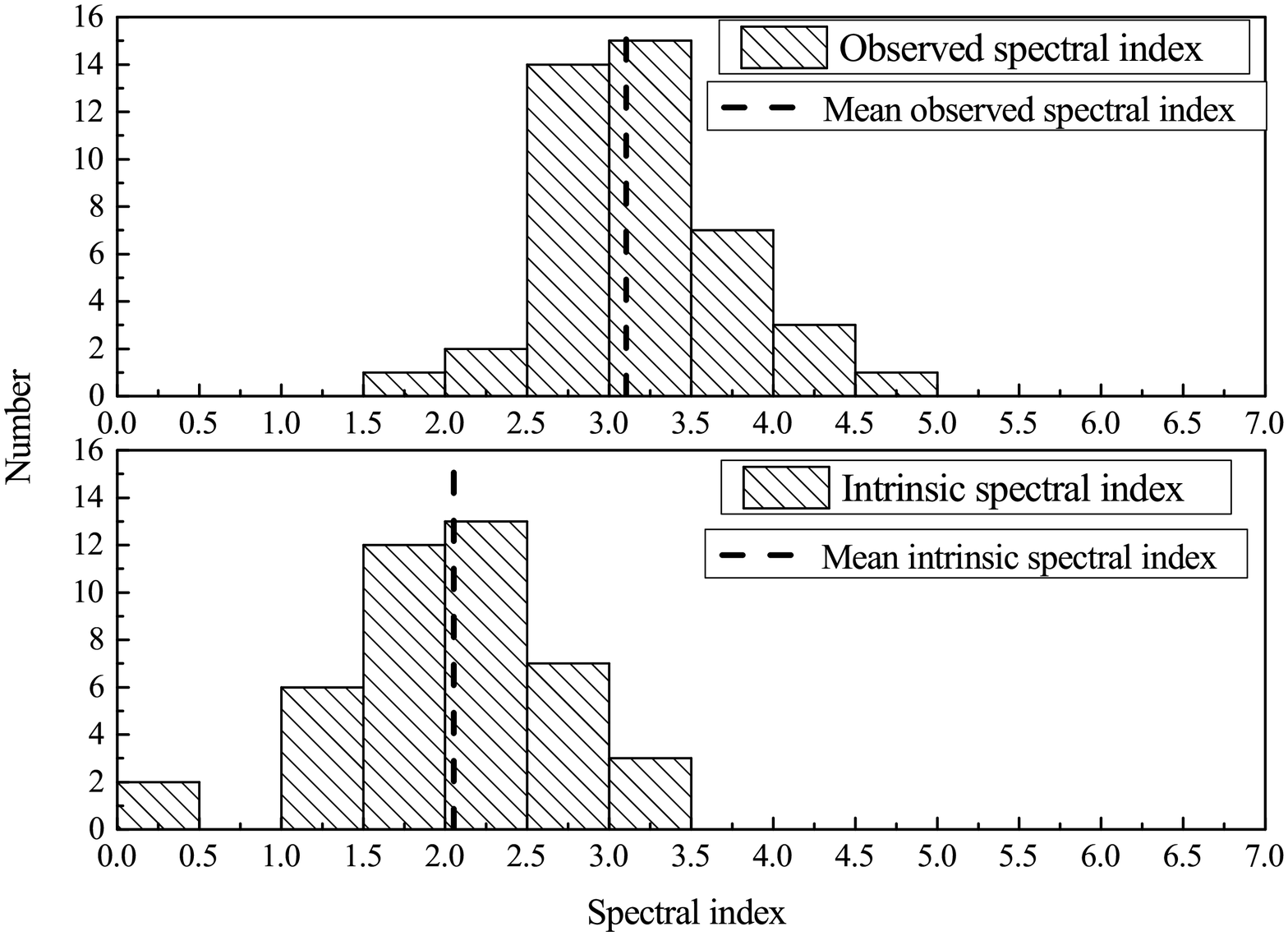}\\
  \caption{Observed spectral indices vs intrinsic spectral indices. As shown in the figure, all of the spectral indices are represented by the black histogram. Black vertical dotted lines represent the mean observed and intrinsic indices that are obtained by the method of \cite{Venters2007The}.}\label{fig:1}
\end{figure}

\begin{figure}[tb]
  \centering
  \includegraphics[width=8cm]{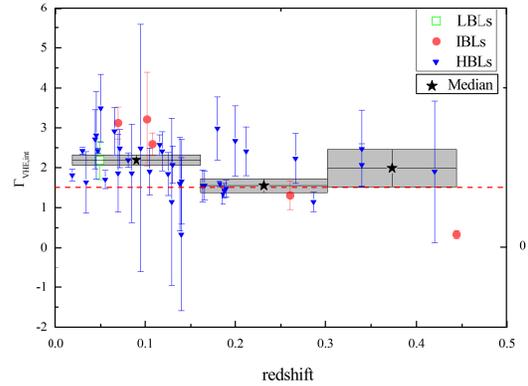}\\
  \caption{Relation of the intrinsic VHE spectral indices with redshift. As shown above, solid red dots represent LBLs, green squares represent IBLs, and solid blue triangles represent HBLs. The error is derived from the MAGIC, HESS, and VERITAS telescopes. Solid black pentagons represent the median of sources within each redshift bin (i.e., we can obtain three redshift bins), the 1$\sigma $ uncertainties of the median are plotted by the gray box. According to our analysis, all our sources are nearly softer than 1.5, and this lower limit is plotted by a red horizontal line.}\label{fig:2}
\end{figure}

The spreads of the observed and intrinsic indices are ${\sigma _{0,obs}} = 0.43$ and ${\sigma _{0,int}} = 1.39$, respectively, and  ${\sigma _{0,obs}}$ is obviously less than ${\sigma _{0,int}}$. In this case, we use a simple mathematical calculation to recalculate their spread, and the results are listed as follows: (1) for all of BL Lacs, ${\sigma _{obs}} = 0.35$ and ${\sigma _{int}} = 0.48$; (2) for HBL objects, ${\sigma _{obs}} = 0.36$ and ${\sigma _{int}} = 0.39$.

We can see that the difference between ${\sigma _{obs}}$ and ${\sigma _{int}}$ for HBL objects is small, while it is large for the case of all BL Lacs, which could also be applied to the case of maximum-likelihood spread. This indicates that the intrinsic property of the synchrotron peak may influence the variance of all BL Lacs. Additionally, the simulation of the intrinsic spectral index with different EBL models will bring some errors to the maximum-likelihood spread. If the above effects are not regarded, the differences between the observed and intrinsic likelihood spreads will be negligible, and it would also imply a similar confidence level for our work.

The redshift evolution of the intrinsic VHE spectral indices is shown in Figure 2. We can see the intrinsic spectral indices do not change with redshift. The median indices (black pentagrams) are in three redshift bins with their $1\sigma$ uncertainties, which are obtained by the distribution of sources in each redshift bin. The median cannot affect the distribution of the VHE intrinsic index. The intrinsic spectral index is obtained by the correction of the EBL, which would be expected to be softer than 1.5 (i.e., ${\Gamma _{{\mathop{\rm int}} }} \ge 1.5$), which directly links the intrinsic spectra with the observed spectra (\citealt{Gilmore2012Semi}, \citealt{Aharonian2006A}). Thus, we can see most of the sources are within 1$\sigma $ and their indices are above the horizontal line graphed with ${\Gamma _{{\mathop{\rm int}} }} = 1.5$, which is typically taken as the hardest spectrum.

\section{Analyzing Spectral Break}
As first suggested by \cite{stecker2006a,Stecker2010Derivation}, we employ the simple linear relation $\Delta\Gamma_{obs}=\alpha\rm{z} + \beta $ for determining the effect of EBL absorption on the observed spectral break. We then use a statistical analysis based on data from the Fermi and TeV catalogs of BL Lac objects to determine the parameters $\alpha$ and $\beta$. Based on the four power-law indices above (i.e., $\Gamma_{VHE, obs}$, $\Gamma_{VHE, int}$,  $\Gamma_{HE, obs}$, and $\Gamma_{HE, int}$), we analyze the difference between the observed and intrinsic indices; that is, the spectral break between the VHEs and HEs. Since the VHE $\gamma$-ray photons could be absorbed by the EBL, we should expect that any significant break in the measured spectrum is the result of EBL absorption (e.g., \citealt{Sanchez2013Evidence}). This results from the relations both $\Gamma_{HE, obs}=\Gamma_{HE, int}$ and $\Gamma_{VHE, obs}=\Gamma_{VHE, int}+\Delta\Gamma_{EBL}(E,z)$. Obviously, the EBL-induced break should increase linearly with the redshift $z$, and we could expect a relationship with $\Delta\Gamma_{EBL}(E,z)=\alpha z$.

We now confirm the relationship between the EBL attenuation and redshift [i.e., $\Delta\Gamma_{EBL}(E,z)=\alpha z$] using the observed results. The EBL attenuation can be represented as the difference between the observed and intrinsic VHE spectral index, e.g., $\Gamma _{VHE,obs} - \Gamma _{VHE,int}$. Figure 3 plots the evolution of the EBL attenuation with redshift (the red dotted line), which is compatible with the result of linear regression; that is, $\Delta {\Gamma _{EBL}(E,z)} =(8.65 \pm 0.13)z + (-0.08 \pm 0.02)$ with a correlation coefficient $r = 0.99$. The influence of EBL attenuation cannot be removed due to $\alpha  = 8.65 \pm 0.13 \ne 0 $, but that from other physical properties that are independent of redshift is negligible due to $\beta  = -0.08 \pm 0.02 \approx 0$. This conclusion satisfies the correlation $\Delta\Gamma_{EBL}(E,z)=\alpha z$.

\begin{figure}[tb]
  \centering
  \includegraphics[width=8cm]{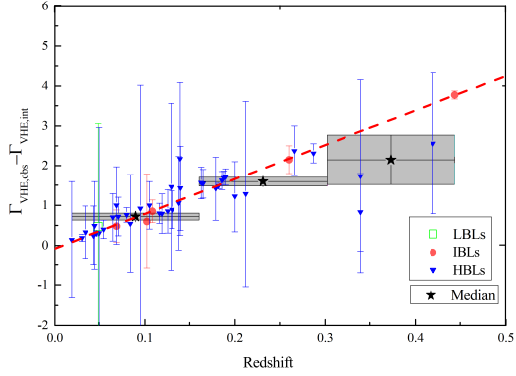}\\
  \caption{Relationship between the EBL attenuation and redshift. According to the equation $\Delta {\Gamma _{EBL}(E,z)} = \Gamma _{VHE,obs} - \Gamma _{VHE,int}$, $\Delta {\Gamma _{EBL}(E,z)} $ reflects the effect of the EBL attenuation. The red dotted line is the best-fit straight line to the data points. Boxes, symbols, and colors of those sources are the same as in Figure 2.}\label{fig:3}
\end{figure}

  The observed spectral break could be represented as the difference between the observed  VHE and HE spectral indices; that is, $\Delta\Gamma_{obs}=\Gamma_{VHE,obs}-\Gamma_{HE,obs}$. Similarly, the intrinsic spectral break could be represented as the difference between the intrinsic VHE and HE spectral index; that is, $\Delta \Gamma_{int}=\Gamma_{VHE,int}-\Gamma_{HE, int}$. Therefore, we can obtain a relation of the observed spectral break and redshift; that is, $\Delta\Gamma_{obs}=\Gamma _{VHE,obs}^{} - \Gamma _{HE,obs}^{} =\Gamma_{VHE,int}-\Gamma_{HE,int}+\Delta\Gamma_{EBL}(E,z)=\Delta\Gamma_{EBL}(E,z)+\Delta \Gamma_{int} =\alpha {\rm{z}} + \beta $, where we expect $\Delta \Gamma_{int} =\beta $ and $\beta$ is related to the intrinsic curvature (\citealt{Stecker2010Derivation}; \citealt{Dom2015Spectral}). Corresponding to the relation of observed spectral break with redshift, we derived the best-fitting line (Figure 4, red dotted line) of the sample and its correlation coefficient, e.g., $\Delta\Gamma_{obs}=(3.60 \pm 0.60) z + (0.83 \pm 0.06)$ and $r = 0.52$.  Since the the expression of the simulation shows that $\alpha  = 3.60 \pm 0.60 \ne 0$ and $\beta  = 0.83\pm 0.06\ne 0$, the observed spectral break is determined by EBL attenuation and intrinsic curvature. This conclusion also indicates why the correlation coefficient in Figure 4 is smaller than that in Figure 3.

\begin{figure}[tb]
  \centering
  \includegraphics[width=8cm]{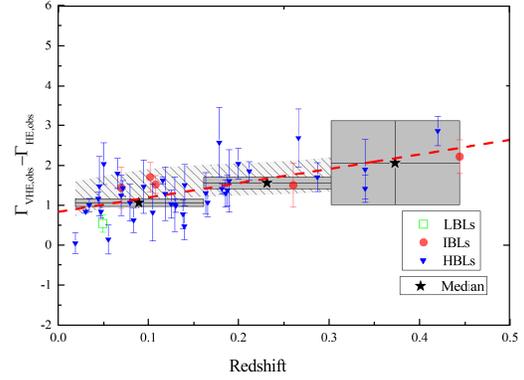}\\
  \caption{Relationship between the observed spectral break and redshift. Errors bars are the sum of the errors of the observed VHE and HE spectral indices, but the boxes, symbols, and colors of the others are the same as in Figure 2. Red dotted line is the best-fit straight line according to the model, $\Delta\Gamma_{obs}=\alpha {\rm{z}} + \beta$. Gray shaded band indicates EBL attenuation and the intrinsic curvature. Note that the signature of spectral break is more obvious at higher redshift. }\label{fig:4}
\end{figure}

 In this mathematical model, the value of $\beta$ relates to the intrinsic curvature that could be represented as the difference between the spectral index before and after the peak frequency(\citealt{Chen2014Curvature}). Based on the above analysis, we must obtain the relationship between the observed spectral break and redshift using theoretical models. Using a synchronous self-Compton (SSC) model with a broken-power-law electron-energy distribution that can account for the intrinsic curvature (\citealt{phdthesis2009}) to simulate the SEDs of all BL Lacs, we calculate the model spectral break (\citealt{Albert2007Discovery} and \citealt{Aharonian2007Discovery}). We show the model results in a $ {\rm{68\% }}$ confidence level as a gray-shadow band in Figure 4. It can be seen that the evolution of the observed spectral break with redshift can be explained by the EBL attenuation.

 The result of the evolution of the intrinsic spectral break (i.e., $\Gamma _{VHE, int} - {\Gamma _{HE,obs}}$) with redshift is shown in Figure 5. The intrinsic spectral break has not suffered from the affect of EBL since the intrinsic spectrum is an EBL unattenuated spectrum. As seen in Figure 5, the intrinsic spectral break is largely independent of redshift, and the distribution of the median also cannot affect this research. Then, we employ the SSC model to calculate the gray-shadow band, and this gray-shadow band is consistent with the relation $\Delta \Gamma_{int} =\beta $. Combined with these methods, we find $\alpha = 0$ and $\beta \ne 0$, indicating that the EBL attenuation cannot affect the intrinsic spectral break, except for some physical processes (intrinsic curvature).
\begin{figure}[tb]
  \centering
  \includegraphics[width=8cm]{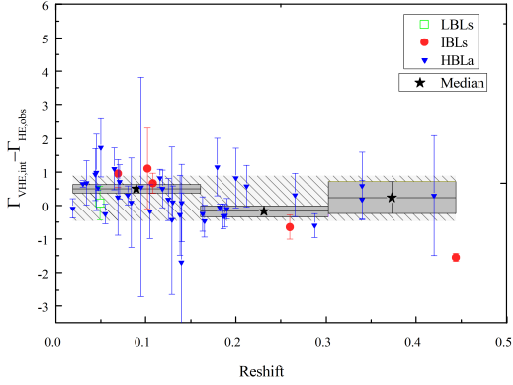}\\
  \caption{Intrinsic spectral break vs redshift. EBL attenuation has been excluded by the property of the intrinsic spectral break; only the blazar physics (intrinsic curvature) can affect this relationship. Gray shaded band is consistent with the mathematical model, $\Delta \Gamma_{int} =\beta $. Boxes, symbols, and colors of those sources are the same as in Figure 2.}\label{fig:5}
\end{figure}

\section{Conclusions and Discussion}
In this paper, we compile the GeV-TeV energy spectral indices of 43 BL Lac objects to analyze the scatter of the $\gamma$-ray observed spectra and intrinsic spectra. We found that the mean observed index is significantly higher than the mean intrinsic index (i.e., $\Gamma _{0, obs}= 3.10$ for observed spectrum and $\Gamma _{0, int}= 2.05 $ for the intrinsic spectrum), implying that the intrinsic spectrum is a hard spectrum. It was also proved that the observed spectrum is attenuated by EBL, which would reduce the mean spectral index.

We focus on a relationship, $\Delta\Gamma_{obs}=\alpha\rm{z} + \beta $, between the observed spectral break and redshift. Three cases exist for this mathematical model, as follows:
\begin{enumerate}%
\item When $\alpha \ne 0$ and $\beta=0$, the equation can be replaced by the relationship between EBL attenuation and redshift, $\Delta {\Gamma _{EBL}(E,z)} = \Gamma _{VHE,obs}- \Gamma _{VHE, int}$, which is not in accord with the observed spectral break. This shows that the EBL attenuation will change with redshift, especially for the high-redshift spectrum.
\item When $\alpha \ne 0$ and $\beta\ne 0$, the origin of the observed spectral break can be affirmed by linear regression and the simulation of theoretical models (i.e., SSC and EBL models). The observed spectral break is determined by the EBL attenuation and the intrinsic curvature (some blazar physics processes) due to the facts that $\alpha \ne 0$ and $\beta=0$.
\item When $\alpha=0 $ and $\beta\ne 0$, the equation relates to the difference between the the intrinsic spectral index at VHEs and spectral index at HEs (intrinsic spectral break), ${\Delta\Gamma _{int}} = {\Gamma _{VHE,int }} - {\Gamma _{HE,obs}}$. In this case, we only study the evolution of the intrinsic spectral break with redshift. The EBL attenuation is removed due to the property of the intrinsic spectral break. From the distribution of intrinsic break and the simulation, the intrinsic curvatures play a crucial role in the intrinsic spectral break, and they cannot evolve with redshift.
\end{enumerate}

 Owing to the statistical results, $\alpha  = 3.60 \pm 0.72 \ne 0$ and $\beta  = 0.83\pm 0.13 \ne 0$, it is suggested that the observed spectral break is dominated by EBL attenuation and the intrinsic curvature (some blazar physics processes). This confirms that EBL attenuation is an origin for the observed spectral break that has evolved with redshift.

Although we have verified that the EBL attenuation is an origin for the observed spectral break of the BL Lac objects, according to the statistical results $\beta \ne 0$ the study of $\beta$ will become indispensable. Some different methods can be used to simulate the spectrum, \cite{Tramacere2011Stochastic} employed the log-parabolic-law (log-parabolic-law electron-energy distribution) SSC model to obtain the intrinsic curvature. However, our research employed the broken-power law SSC model to obtain it. Different methods will lead to different electron spectra, which could affect the intrinsic curvature. Additionally, the intrinsic curvature also can be interpreted by the Klein-Nishina suppression (emission effects) or a turnover in the distribution of the underlying emitting particles (acceleration effects, e.g., \citealt{Sanchez2013Evidence}.) The redshift evolution of the observed spectral break can be explained solely  by EBL attenuation (without the secondary cascades or axion-like particle conversion), and there is also no evidence of evolution with redshift of the physics that drives the photon emission (\citealt{Dom2015Spectral}).

 \acknowledgment
We thank the anonymous referee for valuable comments and suggestions. This work is partially supported by the National Natural Science Foundation of China under grants 11463007, 11673060 and the Natural Science Foundation of Yunnan Province grants 2017FD072.

 \bibliographystyle{spr-mp-nameyear-cnd}  
\bibliography{1r_lamboo_notes}

\begin{deluxetable}{ccrrrrrc}
\tabletypesize{\scriptsize}
\rotate
\tablecaption{BL lac object data.\label{tbl-1}}
\tablehead{
\colhead{Name} & \colhead{Telescope} & \colhead{Type} & \colhead{ Redshift} & \colhead{$\Gamma_{HE, obs}$} & \colhead{$\Gamma_{VHE, obs}$} & \colhead{$\Gamma_{VHE, int}$} & \colhead{Refernce}
}
\startdata
    1ES 0033+595 & MAGIC & HBL   & 0.34  & 1.90 $\pm$ 0.04 & 3.80 $\pm$ 0.70   & 2.07 $\pm$ 0.53  & \cite{Aleksi2014Discovery}\\
    1ES 0229+200 & HESS & HBL   & 0.1396 & 2.03 $\pm$ 0.15 & 2.50  $\pm$ 0.19  & 0.33  $\pm$ 1.92  & \cite{Aharonian2007New} \\
    1ES 0347-121 & HESS  & HBL   & 0.188 & 1.73 $\pm$ 0.16 & 3.10 $\pm$ 0.23  & 1.42  $\pm$ 0.18  & \cite{Aharonian2007Discovery} \\
    1ES 0414+009 & HESS  & HBL   & 0.287 & 1.75 $\pm$ 0.11 & 3.45  $\pm$ 0.25  & 1.14  $\pm$ 0.24  &  \cite{AbramowskiB2012Discovery}\\
    1ES 0806+524 & MAGIC & HBL   & 0.138 & 1.88 $\pm$ 0.02 & 2.65  $\pm$ 0.34  & 1.59  $\pm$ 1.17  &  \cite{Aleksi2016Insights}\\
    1ES 1011+496 & MAGIC & HBL   & 0.212 & 1.83$\pm$ 0.02 & 3.69  $\pm$ 0.22  & 2.41  $\pm$ 0.61  &  \cite{Aleksi2016Insights}\\
    1ES 1101-232 & HESS  & HBL   & 0.186 & 1.64 $\pm$ 0.14 & 2.94  $\pm$ 0.20   & 1.32  $\pm$ 0.23  & \cite{Aharonian2007Detection} \\
    1ES 1215+303 & MAGIC & HBL   & 0.13  & 1.97 $\pm$ 0.02 & 2.96  $\pm$ 0.14  & 2.07  $\pm$ 0.47  &  \cite{Aleksi2012Discovery}\\
    1ES 1218+304 & VERITAS & HBL   & 0.182 & 1.66 $\pm$ 0.04 & 3.07 $\pm$ 0.09  & 1.59  $\pm$ 0.07  &  \cite{Acciarib2010Discovery}\\
    1ES 1312-423 & HESS  & HBL   & 0.105 & 2.08 $\pm$ 0.21 & 2.90 $\pm$ 0.50   & 1.9   $\pm$ 0.59  &  \cite{Abramowski2013HESS}\\
    1ES 1440+122 & VERITAS & HBL   & 0.163 & 1.80 $\pm$ 0.12 & 3.10 $\pm$ 0.40   & 1.54  $\pm$ 0.4   & \cite{Archambault2016Discovery}\\
    1ES 1727+502 & VERITAS & HBL   & 0.055 & 1.96 $\pm$ 0.06 & 2.10 $\pm$ 0.30   & 1.7   $\pm$ 0.23  & \cite{Archambault2015VERITAS}\\
    1ES 1741+196 & MAGIC & HBL   & 0.084 & 1.78 $\pm$ 0.11 & 2.40 $\pm$ 0.20   & 1.86  $\pm$ 1.23  &  \cite{Ahnen2017MAGIC}\\
    1ES 1959+650 & MAGIC & HBL   & 0.047 & 1.88 $\pm$ 0.02 & 2.72 $\pm$ 0.14  & 2.42  $\pm$0.06  &  \cite{AlbertA2006Observation}\\
    1ES 2344+514 & MAGIC & HBL   & 0.044 & 1.78 $\pm$ 0.04 & 2.95  $\pm$ 0.12  & 2.71  $\pm$ 0.74  &  \cite{AlbertJ2007Observation}\\
    1RXS J101015.9 & HESS  & HBL   & 0.14  & 1.58 $\pm$ 0.10 & 3.08 $\pm$ 0.42  & 1.65 $\pm$ 1.06  & \cite{Abramowski2012Discovery} \\
    3C 66A  & VERITAS & IBL   & 0.444 & 1.88 $\pm$ 0.02 & 4.10 $\pm$ 0.40   & 0.33  $\pm$ 0.10   &  \cite{Acciari2009VERITAS}\\
    AP Librae & HESS  & LBL   & 0.049 & 2.11 $\pm$ 0.03 & 2.65 $\pm$ 0.19  & 2.18  $\pm$ 0.48  &  \cite{Abramowski2015The}\\
    B3 2247+381 & MAGIC & HBL   & 0.1187 & 1.91 $\pm$ 0.07 & 3.20 $\pm$ 0.60   & 2.41  $\pm$ 0.49  &  \cite{Aleksi2017Discovery}\\
    BL Lacertae & MAGIC & IBL   & 0.069 & 2.16 $\pm$ 0.017 & 3.60 $\pm$ 0.50   & 3.12  $\pm$ 0.4   &  \cite{AlbertJ2007Observation}\\
    H 1426+428 & HEGRA & HBL   & 0.129 & 1.57 $\pm$ 0.09 & 2.60 $\pm$ 0.60   & 1.14  $\pm$ 2.1   &  \cite{Mueller2011Very}\\
    H 1722+119 & MAGIC & HBL   & 0.34  & 1.89 $\pm$ 0.05 & 3.30 $\pm$ 0.30   & 2.47  $\pm$ 0.97  &  \cite{Ahnen2016Investigating}\\
    H 2356-309 & HESS  & HBL   & 0.165 & 2.02 $\pm$ 0.12 & 3.09 $\pm$ 0.24  & 1.55 $\pm$ 0.36  &  \cite{Aharonian2006Discovery}\\
    IC 310 & MAGIC & HBL   & 0.0189 & 1.90 $\pm$ 0.14 & 1.95 $\pm$ 0.12  & 1.81  $\pm$ 0.15  &  \cite{Aleksi2014Rapid}\\
    Markarian 180 & MAGIC & HBL   & 0.045 & 1.82 $\pm$ 0.05 & 3.30 $\pm$ 0.70   & 2.81  $\pm$ 1.1   & \cite{Albert2006Discovery} \\
    Markarian 421 & MAGIC & HBL   & 0.03  & 1.77 $\pm$0.008 & 2.61  $\pm$ 0.03  & 2.42  $\pm$ 0.09  &  \cite{Acciari2011TeV}\\
    Markarian 501 & VERITAS & HBL   & 0.034 & 1.72 $\pm$ 0.02 & 2.72  $\pm$ 0.15  & 2.39  $\pm$ 0.65  &  TeV Catalogue\\
    PG 1553+113 & HESS  & HBL   & 0.42 & 1.60 $\pm$ 0.03 & 4.46 $\pm$ 0.34  & 1.9   $\pm$ 1.77  &  \cite{Aharonian2007HESS}\\
    PKS 0301-243 & HESS  & HBL   & 0.266 & 1.92 $\pm$ 0.03 & 4.60 $\pm$0.70   & 2.23  $\pm$ 0.63  &  \cite{Abramowski2013Discovery}\\
    PKS 0447-439 & HESS  & HBL   & 0.20   & 1.85 $\pm$ 0.02 & 3.89  $\pm$ 0.37  & 2.67  $\pm$ 0.88  &  \cite{AbramowskiB2013Discovery}\\
    PKS 0548-322 & HESS  & HBL   & 0.069 & 1.61 $\pm$ 0.16 & 2.86  $\pm$ 0.34  & 1.86  $\pm$ 0.97  &  \cite{Aharonian2010Discovery}\\
    PKS 1424+240 & VERITAS & HBL   & 0.05  & 1.76 $\pm$ 0.026 & 3.80 $\pm$ 0.50   & 3.49  $\pm$ 0.85  & \cite{Acciari2010DISCOVERY}\\
    PKS 1440-389 & HESS  & HBL   & 0.065 & 1.81 $\pm$ 0.04 & 3.61 $\pm$ 0.34  & 2.91  $\pm$ 0.59  &  \cite{Prokoph2015H}\\
    PKS 2005-489 & HESS  & HBL   & 0.071 & 1.77 $\pm$ 0.03 & 3.20  $\pm$ 0.16  & 2.48  $\pm$ 0.48  &  \cite{Acero2010PKS}\\
    PKS 2155-304 & HESS  & HBL   & 0.116 & 1.75 $\pm$ 0.02 & 3.37  $\pm$ 0.07  & 2.57  $\pm$ 0.25  &  \cite{Aharonian2005Multi}\\
    RBS 0413 & VERITAS & HBL   & 0.19  & 1.57 $\pm$ 0.10 & 3.18  $\pm$ 0.68  & 1.47 $\pm$ 0.21  &  \cite{Aliu2012DISCOVERY}\\
    RGB J0152+017 & HESS  & HBL   & 0.08  & 1.89 $\pm$ 0.10 & 2.95  $\pm$ 0.36  & 2.19  $\pm$ 0.18  &  \cite{Aharonian2008Discovery}\\
    RGB J0710+591 & VERITAS & HBL   & 0.125 & 1.66 $\pm$ 0.09 & 2.69  $\pm$ 0.26  & 1.84  $\pm$ 0.54  & \cite{Acciari2010The}\\
    RX J0648.7+1516 & VERITAS & HBL   & 0.179 & 1.83 $\pm$ 0.07 & 4.40  $\pm$ 0.80   & 2.98  $\pm$ 0.79  &  \cite{Errando2011Discovery}\\
    S5 0716+714 & MAGIC & IBL   & 0.26  & 1.95 $\pm$ 0.01 & 3.45  $\pm$ 0.54  & 1.3   $\pm$ 0.35  &  \cite{Anderhub2009Discovery}\\
    SHBL J001355.9 & HESS  & HBL   & 0.095 & 1.94$\pm$ 0.17 & 3.4  $\pm$ 0.50   & 2.48  $\pm$ 3.10   & \cite{AbramowskiA2013Discovery}\\
    -185406&&&&&&&\\
    VER J0521+211 & VERITAS & IBL   & 0.108 & 1.92 $\pm$ 0.02 & 3.44 $\pm$ 0.20   & 2.59  $\pm$ 0.28  & \cite{Archambault2013Discovery}\\
    W Comae & VERITAS & IBL   & 0.102 & 2.10$\pm$ 0.03 & 3.81  $\pm$ 0.35  & 3.21  $\pm$ 1.18  &  \cite{Acciari2008VERITAS}\\

\enddata
\tablecomments{The fifth column contains HE observed spectral indices from the 3FGL Catalog. The sixth column contains the VHE observed spectral indices, which are related to the data in the eighth column. The redshift of some sources cannot be ensured, but they are assumed by the references in the eighth column. The seventh column contains the VHE intrinsic spectral break, which we obtain using the average EBL model (\citealt{Dwek2005Simultaneous})}.
\end{deluxetable}

\end{document}